\documentclass[11pt]{article}
\usepackage{epsfig}
\usepackage{graphicx}
\usepackage{leftidx}
\def \be{\begin{equation}}
\def \ee{\end{equation}}
\def \bea{\begin{equnarray}}
\def \eea{\end{eqnarray}}
\begin{document}

\title{\bf \large Study of Single $W$ production in $e\gamma$ collisions through the
decay lepton spectrum to probe $\gamma WW$ couplings}
\vskip 1cm
\author{\bf Satendra Kumar and P. Poulose \\[2mm]
Department of Physics\\
Indian Institute of Technology Guwahati\\
Assam 781 039, India}
\maketitle
\begin{abstract}
We investigate the effect of anomalous $WW\gamma$ couplings
in the $e\gamma\rightarrow \nu W$ through the angular and energy spectrum
of the secondary leptons. Within the narrow-width approximation, a semi-analytical 
study of the secondary lepton energy-angle double distribution is considered.
Utility of observables derived from this is demonstrated by considering the 
anomalous couplings, $\delta \kappa_\gamma$. Results of our
investigation for typical ILC machine considered at $\sqrt{s}=300-1000$ GeV 
re-affirms potential of this collider as a precision machine.
\end{abstract}

\newpage 
\section{Introduction}
While it is true that the Standard Model (SM) has been extensively tested very 
successfully by many different experiments, it is widely believed that
the SM is an effective theory which needs to incorporate suitable modifications 
at large energies. Many expect that this large energy scale could be as small as 
a few TeV, which is right at the verge of being explored. Among the list of less rigorously
tested aspects of the SM tops the mechanism of electroweak symmetry breaking (EWSB) and
the structure and values of gauge-boson self interactions. Explorations of LEP has largely
constrained these areas, but left a lot to future investigations. 
The Large Hadron Collider (LHC) promises a near complete study of the Higgs mechanism of the SM, and perhaps will
bring out information on other mechanisms of EWSB as well. At the same time, being a hadronic
machine it has limitations to undertake precision studies, for example the kind needed
to understand the gauge-boson couplings. The International Linear Collider (ILC) 
proposed to collide high energy, high luminosity electrons and positrons has 
the mission of studying the standard model (SM) at high precision
and to look for signals beyond the standard model~\cite{ILC}. Such a machine is well
suited to an in-depth analysis of the gauge sector of elementary particle interactions,
within and beyond the SM. A large number of physics studies establish the fact that 
the potential of ILC is further enhanced by considering high energy photon-photon 
collisions as well as electron-photon collisions, apart from the electron-positron 
collisions. Technical feasibility of such options are already 
studied in detail, and it is now expected that ILC will have these options available.

Obviously, the $e\gamma$ and $\gamma\gamma$ colliders are better suited
to study the photon couplings with other gauge bosons like the $\gamma WW$, $\gamma\gamma WW$,
$\gamma ZZ$ and $\gamma\gamma Z$. In this article we will focus on 
$e\gamma\rightarrow \nu W$ with $W\rightarrow l\bar\nu$. Signature of such an event is a single isolated lepton with large missing energy. This process is sensitive to new physics effects including anomalous   $\gamma WW$ \cite{pheno:egamWnu, pheno:Godfrey}, composite fermion models \cite{pheno:Gregores}, etc. In the case of anomalous $\gamma WW$ couplings, 
this process has the advantage over $e^+e^-\rightarrow W^+W^-$, which is sensitive
to both $\gamma WW$ and $WWZ$ couplings. Again, $\gamma\gamma\rightarrow W^+W^-$ is 
sensitive to $\gamma WW$, $\gamma\gamma WW$ and $\gamma\gamma Z$ couplings, revealing the edge of $e\gamma$ collider to study $\gamma WW$. In most of the previous studies, observables at the production level of the single $W$ is investigated, with the exception of \cite{pheno:Godfrey}, where the authors consider angular spectrum of the secondary leptons, also including the effect of off-shell $W$. Our main aim of this work is to study the possibility to exploit the secondary lepton spectrum including the energy and angular distributions to probe relevant new physics signals. 
To illustrate the idea, we consider the case of anomalous $\gamma WW$ coupling.
 
In many models beyond the SM the quartic- and triple-gauge boson couplings including 
$\gamma WW$are altered from their SM values. In a model independent approach, an effective 
Lagrangian with terms additional to the SM Lagrangian is considered in phenomenological 
and experimental studies \cite{Hagiwara}.  Relevant to the process considered here, the effective $\gamma WW$ vertex is commonly parametrized in terms of $\delta \kappa_\gamma$ and $\lambda_\gamma$, in the absence of CP violation (with vanishing SM values).  LEP constraints in single-parameter analysis (taking one parameter at a time, keeping the others at their SM values) gives bounds of $-0.105<\delta \kappa_\gamma<+0.069$  and $-0.059<\lambda_\gamma<+0.026$, and two-parameter analysis limits their values to $-0.072<\delta \kappa_\gamma<+0.127$ and $-0.068<\lambda_\gamma<+0.023$ \cite{LEPconstraints} at 95\% C.L. Tevatron constraints from $W\gamma$ process 
are not contaminated by other couplings, but are more relaxed compared the LEP constraints to give $-0.51<\delta \kappa_\gamma<+0.51$ and $-0.12<\lambda_\gamma<+0.13$ at 95\% C.L.

Phenomenology of anomalous $\gamma WW$ coupling in the context of LHC as well as ILC has been carried out in a number of recent publications \cite{pheno:egamWnu, pheno:Godfrey, LHCstudies, ILCstudies}.
 In particular, \cite{pheno:Godfrey} has analyzed single $W$ production with its leptonic decay to probe the effect of anomalous couplings in $e\gamma$ collision. In this work we exploit the full potential of the secondary lepton spectrum to study the effect of $\gamma WW$ coupling in $e\gamma \rightarrow \nu W \rightarrow \nu (l\bar \nu)$ with $\delta \kappa_\gamma$ deviating from its SM value. We leave out the dimension six operator corresponding to $\lambda_\gamma$ from this analysis.
 
 In the next section we provide some details of the process and the observables used. In Section 3 we present our numerical results, and finally summarize the study and present our conclusions in the last section.

\section{Analysis and discussion}

Considering a real on-shell photon, the most general CP-conserving $\gamma WW$ coupling within a  Lorentz invariant Lagrangian can be written in the following form \cite{Hagiwara}.
\begin{eqnarray}
{\cal L}_{\gamma WW}=-ie~\left\{ W^\dagger_{\mu\nu}W^\mu A^\nu-W^\dagger \mu A_\nu W^{\mu\nu}\right.&+&\left. (1+\delta \kappa_\gamma)~W^\dagger_\mu W_\nu F^{\mu\nu}\right.\nonumber \\
&+&\left.\frac{\lambda_\gamma}
{m_W^2}~W^\dagger_{\lambda\mu}W^\mu_{~~\nu}~F^{\lambda\nu}\right\}
\label{eqn:Leff}
\end{eqnarray}
 
\begin{figure}[htbp]
\begin{center}
\epsfig{file=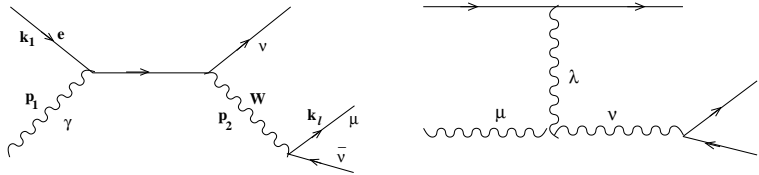,width=10cm, angle=0 }
\caption{Feynman diagrams for the process $e\gamma\rightarrow \nu W\rightarrow \nu (\mu\bar \nu)$. First diagram shows the momenta used and second diagram shows the indices used in the vertex
in Eq.~\ref{eqn:Vertex}}
\label{fig:Feyn}
\end{center}
\end{figure}

 In the SM, the gauge structure $SU(2)_L\times U(1)_Y$ dictates the $\gamma WW$ couplings, with vanishing $\delta \kappa_\gamma$ and $\lambda_\gamma$ at tree level.  Therefore, precise measurements of these couplings will test the gauge sector of the electroweak interactions. Fig.~\ref{fig:Feyn} shows the Feynman diagrams for the process along with the momentum labels.
 
With the effective Lagrangian given by Eq.~\ref{eqn:Leff}, the $\gamma WW$ vertex for the process under study (Fig.~\ref{fig:Feyn} $(b)$)  takes the form:
\begin{eqnarray}
\Gamma^{\mu\nu\lambda}&=&ie~\left\{  
2p_2^\mu~g^{\nu\lambda}+2p_1^\nu~g^{\mu\lambda}-(p_1+p_2)^\lambda~g^{\mu\nu}\right.
\nonumber\\
&&\left. +
\left( \delta \kappa_\gamma-\lambda_\gamma\right)~\left(p_1^\nu~g^{\mu\lambda} -
p_1^\lambda~g^{\mu\nu}\right)\right.\nonumber\\
&& \left.+\frac{\lambda_\gamma}{m_W^2}~
\left(p_1+p_2\right)^\lambda~\left( p_2^\mu p_1^\nu-\left(p_1\cdot p_2\right)~g^{\mu\nu}  \right)
  \right\}
  \label{eqn:Vertex}
\end{eqnarray}

The effective Lagrangian in Eq.~\ref{eqn:Leff} should be considered as a low energy approximation of some fundamental theory, which is expected to emerge at some high energy scale, $\Lambda$. 
To control unitarity violation at high energies, we consider the anomalous couplings as form factors according to the following \cite{pheno:Godfrey}
\begin{equation}
A = A_0~\left[ \left(1+\frac{|p_2^2|}{\Lambda^2}\right)
\left(1+\frac{|(p_1-p_2)^2|}{\Lambda^2}\right) \right],
\label{eqn:HEcompletion}
\end{equation}
where $A\equiv \delta \kappa_\gamma,~\lambda_\gamma$. 

As discussed in the introduction, the main aim of this work is to explore possibilities to exploit energy and angle distributions of the secondary muon to probe the physics beyond the SM. We aim to exemplify this with a study of the anomalous $\gamma WW$ coupling. In what follows, we will therefore limit our analysis to $\lambda_\gamma=0$. Part of the reason is to add clarity to the main focus of the study presented. A more complete study of effect of anomalous couplings in $e\gamma$ collision will be carried out in a future work.

We perform our computation in the Narrow-Width Approximation (NWA) in which the $W$-propagator is approximated to get 
\begin{equation}
\frac{1}{|p_2^2-m_W^2|^2}=\frac{\pi}{m_W\Gamma_W}~\delta(p_2^2-m_W^2),
\label{eqn:NWA}
\end{equation}
 where $m_W$ is the mass and $\Gamma_W$ is the width of the $W$ boson.
 \begin{figure}[htbp]
\begin{center}
\epsfig{file=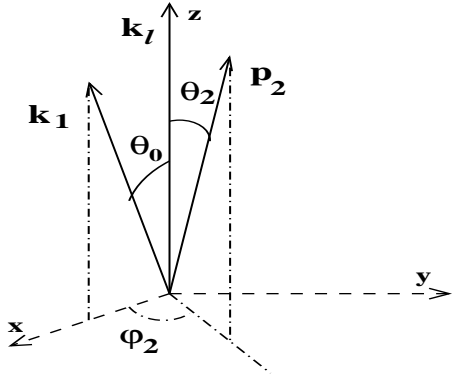,width=6cm, height=5cm,angle=0 }
\caption{Reference frame defining different angles used in Eq.~\ref{eqn:ddist0}.}
\label{fig:RF}
\end{center}
\end{figure}

To perform phase space integrations we fix our reference frame as the centre of mass frame (CMF) of the colliding electron and photon system. $z$-axis is taken along $\vec{k_l}$, which is the momentum of the outgoing lepton (considered as muon in the further discussion) as pictured in Fig.~\ref{fig:RF}. $y$-axis is defined as $\vec{k_l}\times \vec{k_1}$, where $\vec{k_1}$ is the momentum of the colliding electron. The $W$ comes out at a polar angle $\theta_2$ and azymuthal  angle $\phi_2$. Energy-momentum conservation and the NWA (Eq.~\ref{eqn:NWA}) are used to get the differential cross section
\begin{equation}
\frac{d\sigma}{dE_l~d\cos\theta_0~d\phi_2}=\frac{1}{(2\pi)^3}\frac{m_W}{16\Gamma_W}
\frac{1}{E_l(\hat s+m_W^2-2\sqrt{\hat s}E_l)^2}~|M_r|^2,
\label{eqn:ddist0}
\end{equation}
where $\sqrt{\hat s}$ is the center of mass energy, $E_l$ is the energy of the muon and $\cos\theta_0= \frac{\vec{k_l}\cdot \vec{k_1}}{|\vec{k_l}\cdot \vec{k_1}|}$. 
Here $M_r$ is the reduced amplitude given in terms of the invariant amplitude $M$ as,
\begin{equation}
M = \frac{1}{(p_2^2-m_W^2)}~M_r.
\end{equation}
$|M_r|^2$ is obtained using FORM computational package \cite{FORM}. After integrating the uninteresting  $\phi_2$, we get the double distribution of energy and polar angle of the secondary muon in the centre of mass frame of the colliding particles with the electron momentum now taken along the redefined $z$ axis. Notice that the muon energy in the CMF is bounded by $\frac{m_W^2}{2\sqrt{\hat s}} \le  E_l \le \frac{\sqrt{\hat s}}{2}$.

To obtain the distribution in the lab frame, we need to boost the above differential cross section appropriately. For an electron beam energy of $E_e$ and photon energy of $\omega_\gamma=x~E_e$, we have the following relation between the variables in the CMF and the laboratory frame.
\begin{eqnarray}
\hat s &= &x~4E_e^2\nonumber\\
E_l&=&E_l^{lab}~\gamma~\left(1-\beta~\cos\theta^{lab}\right) \nonumber \\
\cos\theta_0&=&\frac{\cos\theta^{lab}-\beta}{1-\beta~\cos\theta^{lab}},
\label{eqn:lab2cmf}
\end{eqnarray}
where $\beta$ is the speed of the CMF compared to the lab frame, and $\gamma=\frac{1}{\sqrt{1-\beta^2}}$. Notice that the limits of $E_l^{lab}$ integration depends on $\cos\theta^{lab}$, keeping it within the bound 
\begin{equation}
\frac{m_W^2}{4\sqrt{x}E_e} \frac{1}{\gamma \left(
1-\beta \cos\theta^{lab}\right)}\le  E_l^{lab} \le\frac{ \sqrt{x}E_e}{\gamma \left(
1-\beta \cos\theta^{lab}\right)}.
\label{eqn:boundsElab}
\end{equation}

 Finally we need to fold the cross section with the appropriate photon distribution function, $f_{\gamma/e}(x)$, an expression for which is provided in the Appendix, so that the total cross section in the lab frame is given by
 \begin{equation}
\sigma = \int f_{\gamma/e}(x)~\sigma(\hat s)~dx.
\label{eqn:lumfold}
\end{equation}

We use Eqs.~\ref{eqn:ddist0}, \ref{eqn:lab2cmf}, \ref{eqn:lumfold} to obtain the total cross section, the muon angular distribution $\frac{d\sigma}{d\cos\theta^{lab}}$, the scaled muon energy distribution $\frac{d\sigma}{dx_l}$, and the energy-angle distribution of the muons$
\frac{d\sigma}{dx_l~d\cos\theta^{lab}}
$ in the lab frame.
For convenience we have defined the dimensionless variable, $x_l=\frac{E_l^{lab}}{E_e}$ to represent the scaled muon energy in the lab. Integrations over the photon distribution variable $x$, the scaled muon energy $x_l$ and the muon angular variable $\cos\theta^{lab}$ in appropriate cases are performed numerically using the {\tt Cuhre} routines under the {\tt CUBA} package 
\cite{CUBA}. From now on we will drop the superscript {\small $lab$} from the variables, and interchangeably use $E_\mu$ or $E_l$ to denote the secondary lepton (muon) energy in the lab frame.

Phenomenological analysis of $e\gamma\rightarrow \nu W$ is considered by some authors in the past \cite{pheno:egamWnu, pheno:Godfrey}. Most of the studies limit their analysis at the production level. Experimentally it is more useful to understand the effect on the final state particles arising from $W$ decay. This is all the more important in the case of leptonic decay, as it is not possible to reconstruct the $W$ in such case.  In ref.~\cite{pheno:Godfrey}, analysis including decay spectrum is presented, where detailed study of the secondary lepton angular distribution is considered along with other reconstructed observables concerning the $W$ production.  In the analysis presented here we demonstrate the usefulness of the combined energy-angle distribution of the secondary leptons in extracting information on the anomalous $\gamma WW$ couplings. While ref.~\cite{pheno:Godfrey} takes into account the contribution due
to off-shell $W$ along with the on-shell production, our work is in the NWA assuming on-shell $W$ production.  At the same time, the semi-analytical method presented gives a handle on the application of useful experimental cuts, as we demonstrate in the next section.

\section{Numerical Results}

For our numerical analysis we consider an ILC, with the option of $e\gamma$ collision using 
backscattered laser photons, expected to run from  $e^+e^-$ (or $e^-e^-$) centre of mass energy of 300 GeV through 1 TeV, with possible extension to higher energies.  
When we discuss observables at specific centre of mass energy values, we will take two example values of centre of mass energy of  $500$ GeV and $1$ TeV.  

For the anomalous couplings, we consider the boundary points from the LEP results, 
\begin{equation}
\delta \kappa_\gamma= -0.072,~+0.069.
\end{equation}

\begin{figure}[htbp]
\begin{center}
\epsfig{file=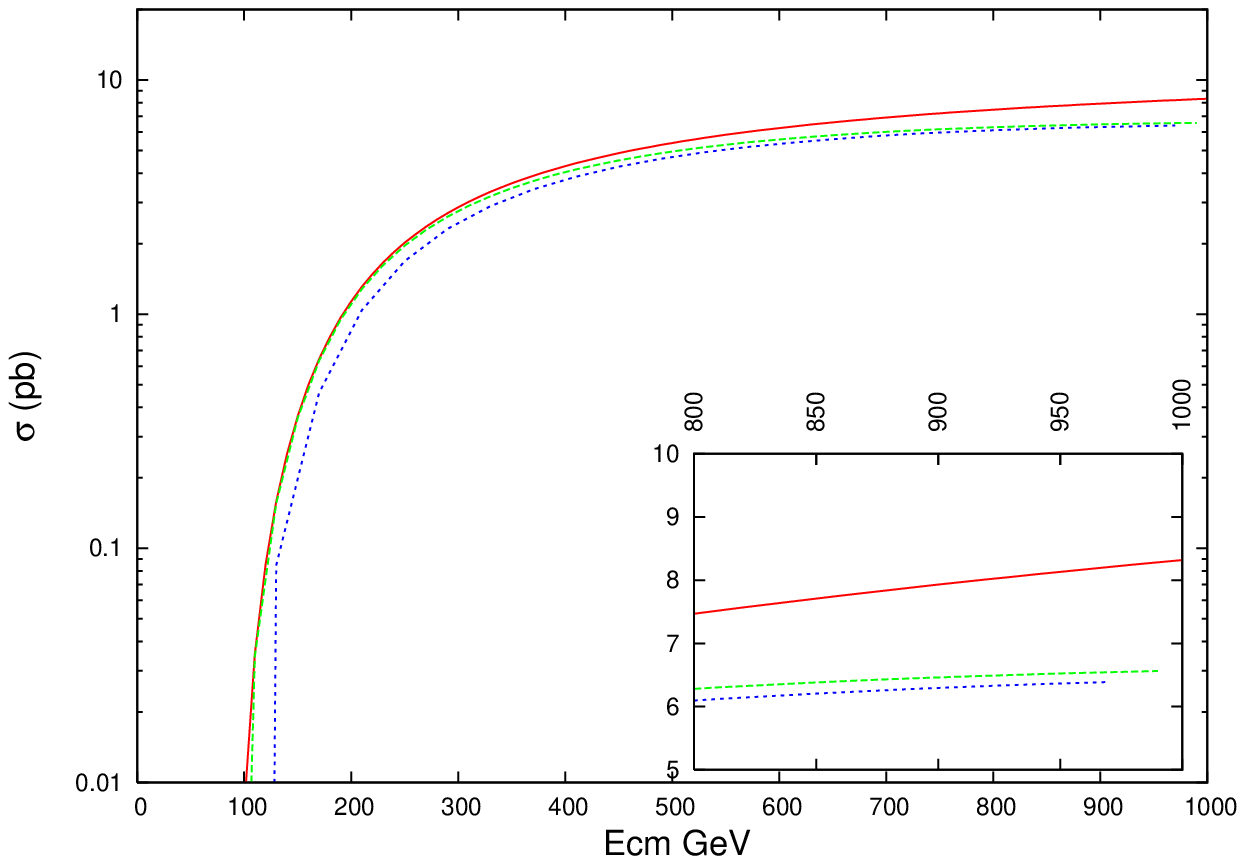,width=6cm, angle=0}
\epsfig{file=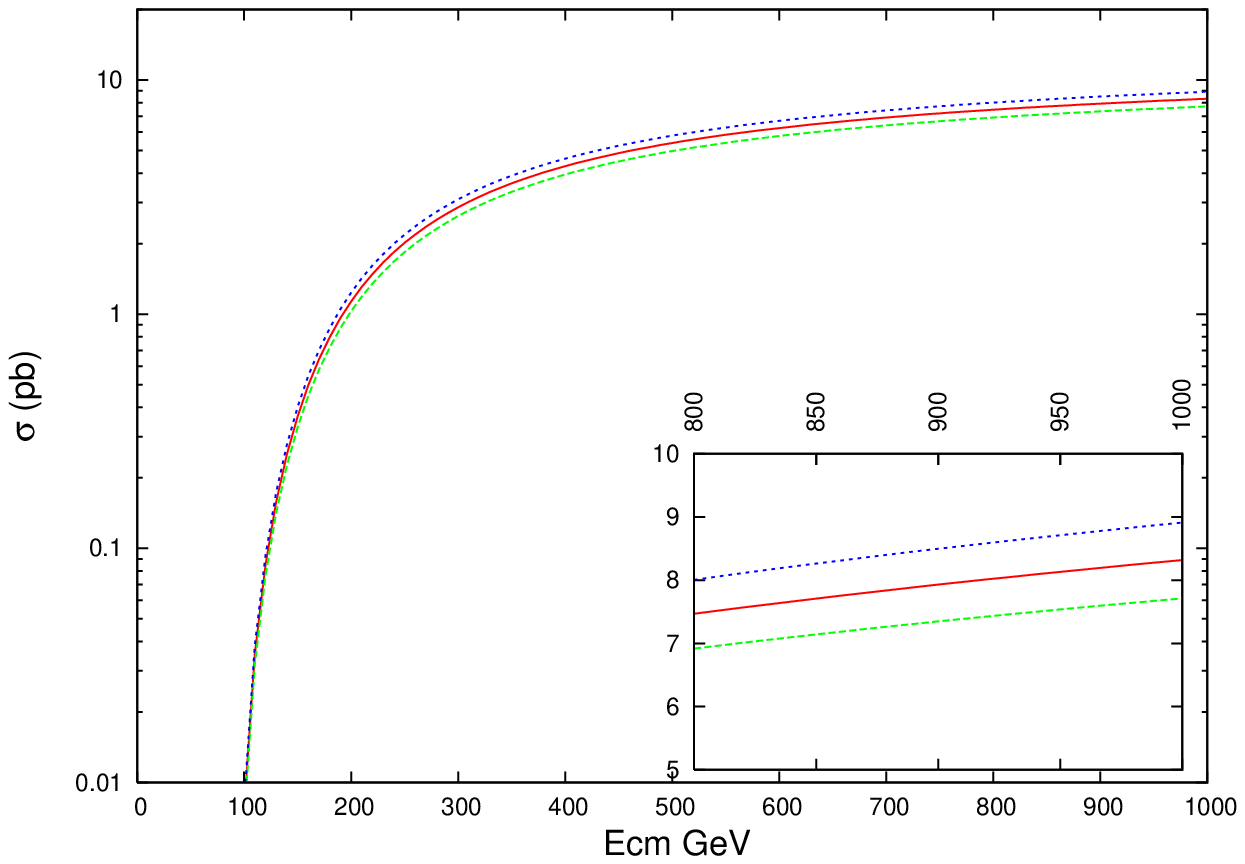,width=6cm, angle=0 }
\caption{{ Total cross section against $E_{cm}=2E_e$. 
{\it Left:} SM cross section with (i) no cuts - (red) solid line, (ii) $10^o\le\theta\le170^o$ - (green) dashed and (iii)
$10^o\le\theta\le170^o~~\&~~E_\mu\ge100$ GeV - (blue) dotted. {\it Right:} 
 without cuts (i) SM: (red) solid line (ii) $\delta\kappa_\gamma=-0.072$ : (green) lower curve and (iii) $\delta\kappa_\gamma=+0.069$ : (blue) upper curve.
 }}
\label{fig:sigvsroots}
\end{center}
\end{figure}

For completeness, we first consider the total cross section. Fig.~\ref{fig:sigvsroots} presents the total cross section against the centre of mass energy of the $e^-e^-$ system (denoted as $E_{cm}$), one of which Compton-scatters on the laser beam to produce the high energy $\gamma$ beam. Figure on the left corresponds SM showing the effect of muon energy ($E_\mu$) and muon angle ($\theta$) cuts. As is seen, angular cut is somewhat sensitive, but the cut on $E_\mu$ has no significant 
effect. Reasons for this will be clear when we discuss the distributions below. In figure on the right, the total cross section in the presence of anomalous coupling is compared with that of the SM case.   
The sensitivity is slightly less than 10\% for most of the $E_{cm}$ beyond $\sim 200$ GeV.

\begin{figure}[htbp]
\begin{center}
\epsfig{file=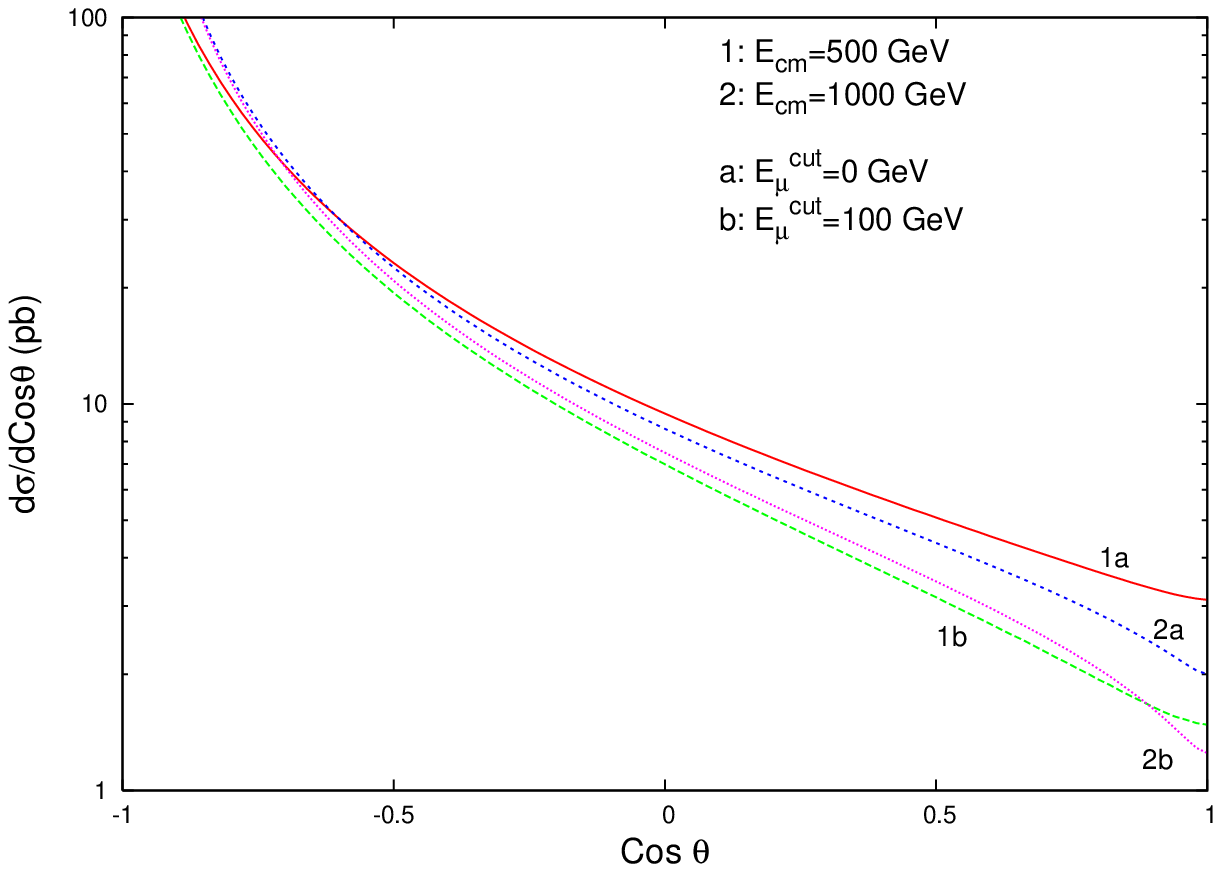,width=6cm, angle=0 }
\epsfig{file=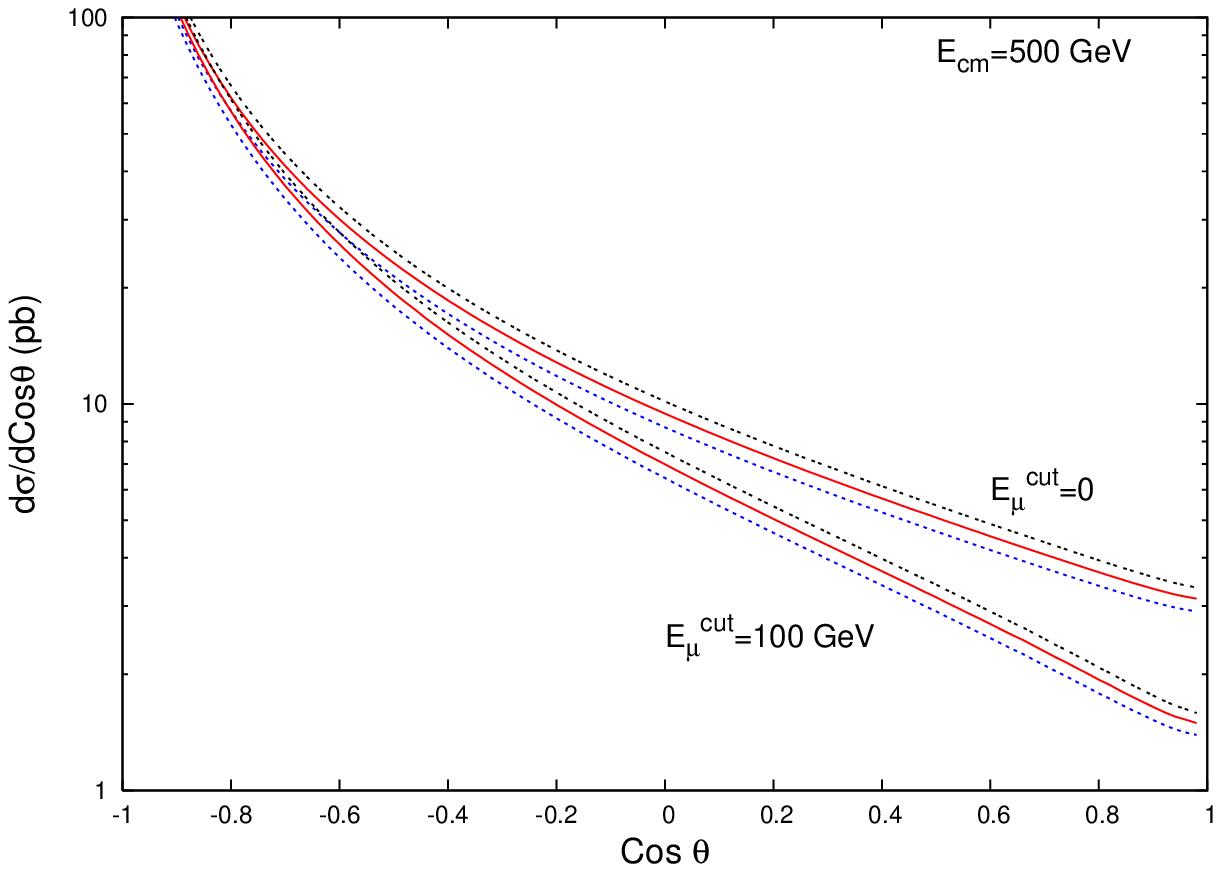,width=6cm, angle=0 }
\caption{{ Angular distribution of the secondary muon in the lab frame at specific values of  $E_{cm}$. {\it Left:} SM case with and without $E_\mu$ cut at two different values of $E_{cm}=500$ GeV and $1$ TeV. {\it Right:} At $E_{cm}=500$ GeV the case of anomalous couplings with
 $\delta\kappa_\gamma=-0.072$ (blue-dotted) and $\delta\kappa_\gamma=+0.069$ (black-dashed) are compared with the case of SM (red - solid) showing the effect of cut on $E_\mu$.
}}
\label{fig:dsigct}
\end{center}
\end{figure}

We next consider the angular distribution of the secondary muons, which is plotted in Fig.~\ref{fig:dsigct}. Figure on the left shows the effect of cut on $E_\mu$ on the SM distributions at centre of mass energies of 500 GeV and 1 TeV. As can be seen, the effect is more pronounced at larger $\cos\theta$ values, while most of the events are gathered in the backward direction. With high luminosity, one expect high statistics for single $W$ production at $e\gamma$ collider. In such case the tail region could also perhaps be probed. In the right side figure angular distribution at $E_{cm}=500$ GeV is analyzed for effects of anomalous couplings. The shape remains the same in all cases, when we add anomalous couplings to the SM case. The effect is not very significant, and neither does the $E_\mu$ cut seem to help.

\begin{figure}[htbp]
\begin{center}
\epsfig{file=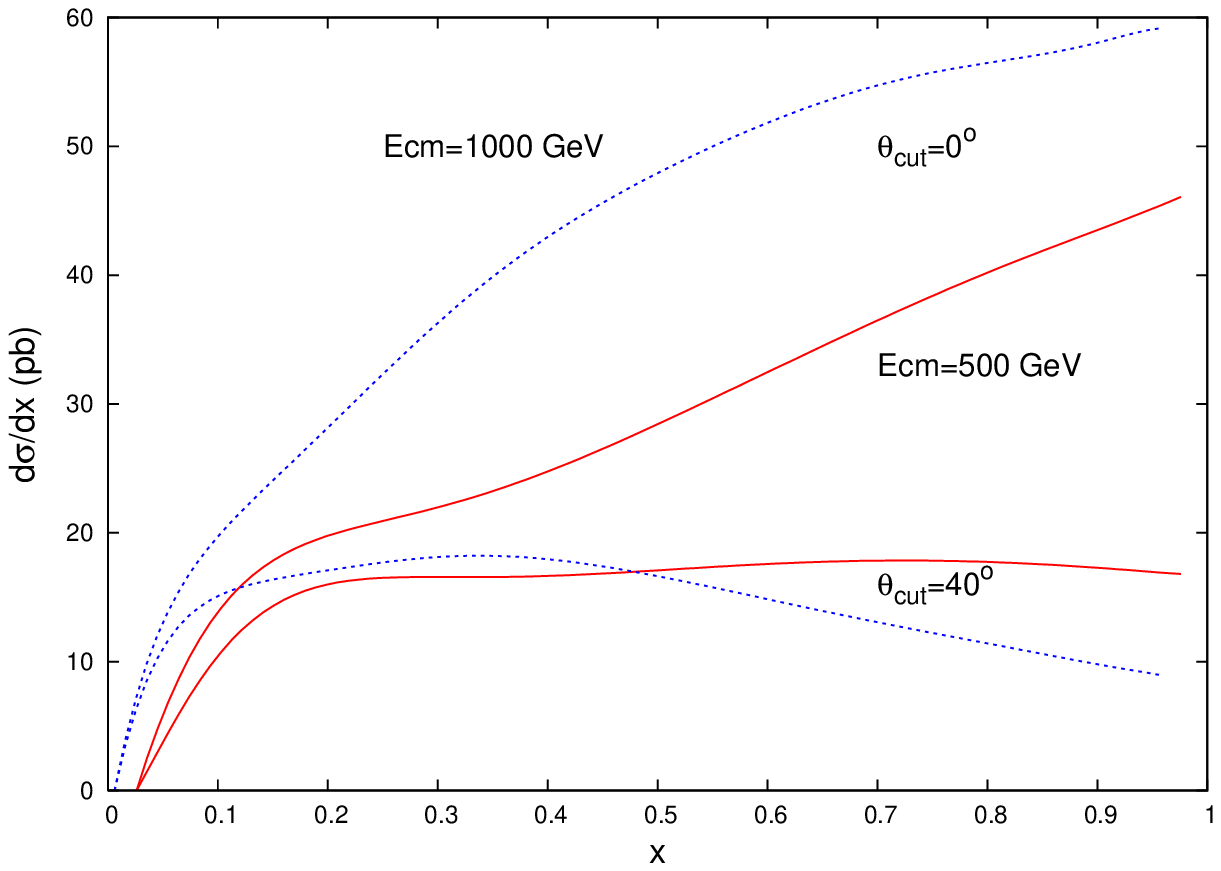,width=6cm, angle=0 }
\epsfig{file=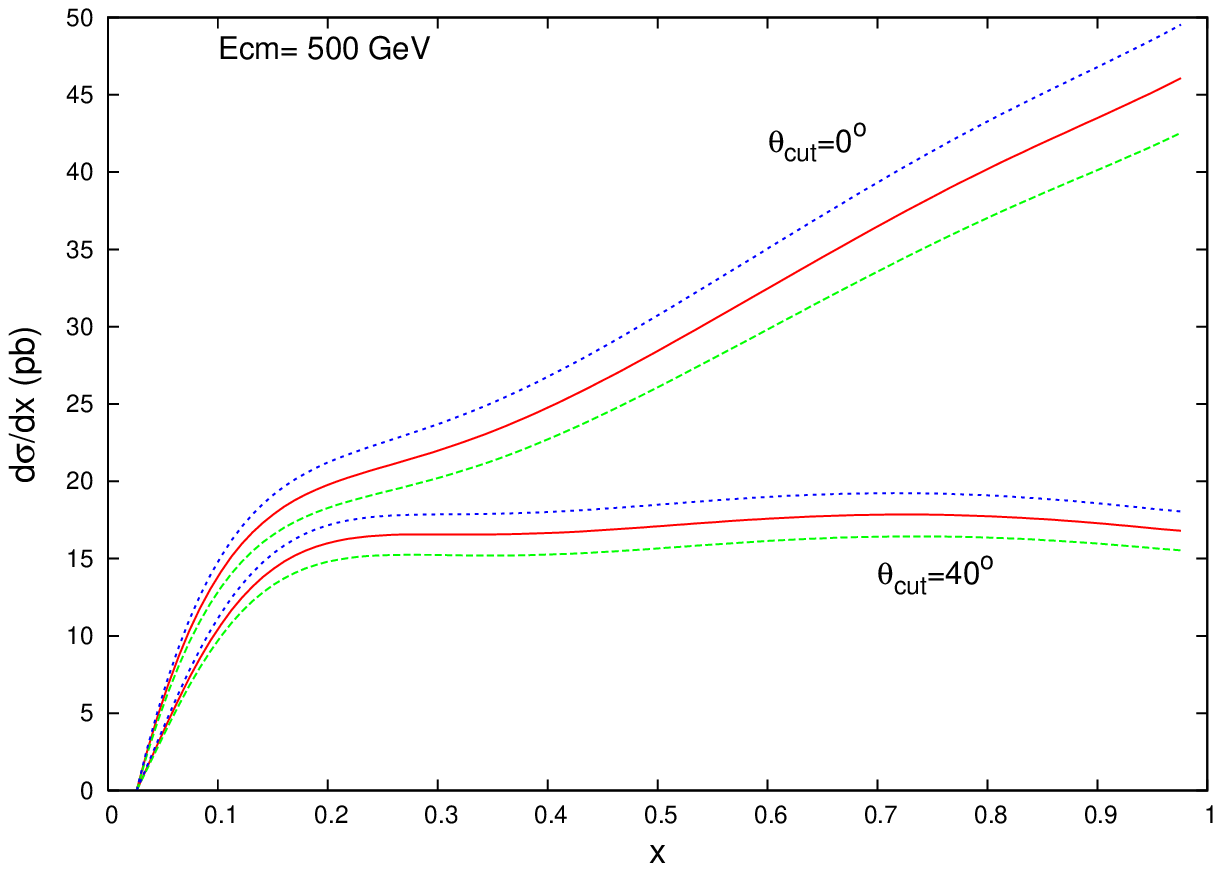,width=6cm}
\caption{{
Energy distribution of the secondary muon in the lab frame at specific values of  $E_{cm}$. {\it Left:} SM case with and without cut on muon angle at two different values of $E_{cm}=500$ GeV and $1$ TeV. {\it Right:} At $E_{cm}=500$ GeV the case of anomalous couplings with
 $\delta\kappa_\gamma=-0.072$ (green-dashed) and $\delta\kappa_\gamma=+0.069$ (blue -dotted) are compared with the case of SM (red - solid) showing the effect of cut on $\theta$.
}}
\label{fig:dsigX}
\end{center}
\end{figure}

To understand the energy spectrum of the muons produced, in Fig.~\ref{fig:dsigX} we plot the distribution against $x=\frac{E_\mu}{E_e}$, where $E_e=\frac{E_{cm}}{2}$ is the electron beam energy.  Notice that the distribution is qualitatively different for different $E_{cm}$. Also the effect of cut on $\theta$ is larger for higher $E_{cm}$. This show that for higher center of mass energies, the events are more clustered along the beam.  Figure on the right side shows the effect of anomalous couplings in the distribution.  Here, the sensitivity is marginally improved at higher $E_\mu$ values compared to lower values. When no angular cut is applied, more than 10\% deviation is observed for both the values of $\delta \kappa_\gamma$ used. The effect is slightly better at larger centre of mass energy values. Angular cut brings down both statistics as well as sensitivity. Perhaps an asymmetric angular cut is better in this case.

\begin{figure}[htbp]
\begin{center}
\epsfig{file=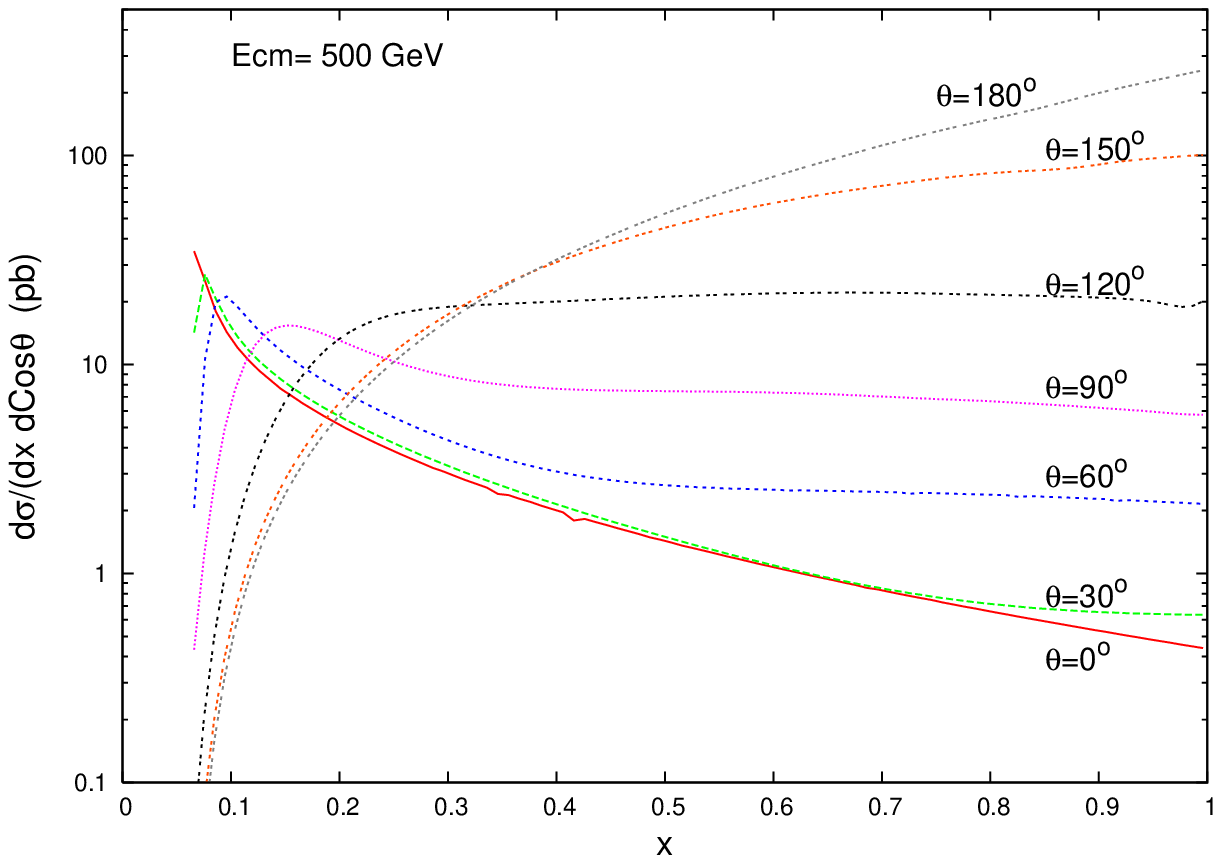,   width=6cm, angle=0}
\epsfig{file=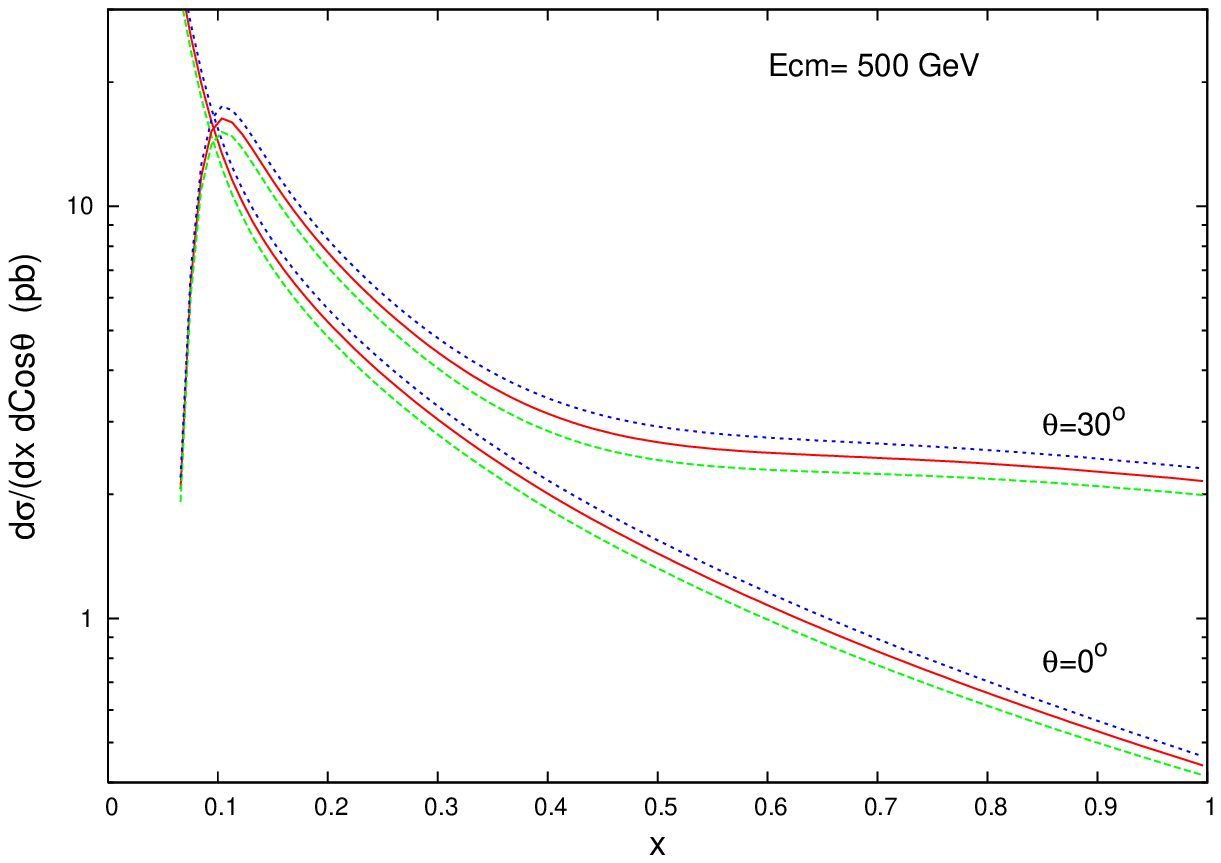, width=6cm, angle=0}
\caption{{
Energy distribution of the secondary muon in the lab frame at specific values of  $E_{cm}$ collected at a particular angle. {\it Left:} SM case at  $E_{cm}=500$ GeV  for different values of $\theta$. {\it Right:} At $E_{cm}=500$ GeV and at two different values of $\theta$, the case of anomalous couplings with
 $\delta\kappa_\gamma=-0.072$ (green-dashed) and $\delta\kappa_\gamma=+0.069$ (blue-dotted) are compared with the case of SM (red - solid).
}}
\label{fig:dsigXCt}
\end{center}
\end{figure}

\begin{figure}[htbp]
\begin{center}
\epsfig{file=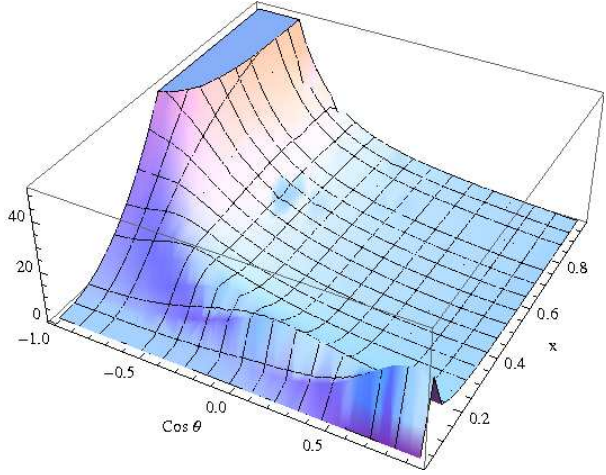, width=8cm, angle=0}
\caption{{3D plot showing energy-angle distribution of secondary muons in the case of SM at $E_{cm}=500$ GeV.
}}
\label{fig:xct3D}
\end{center}
\end{figure}

\begin{figure}[htbp]
\begin{center}
\epsfig{file=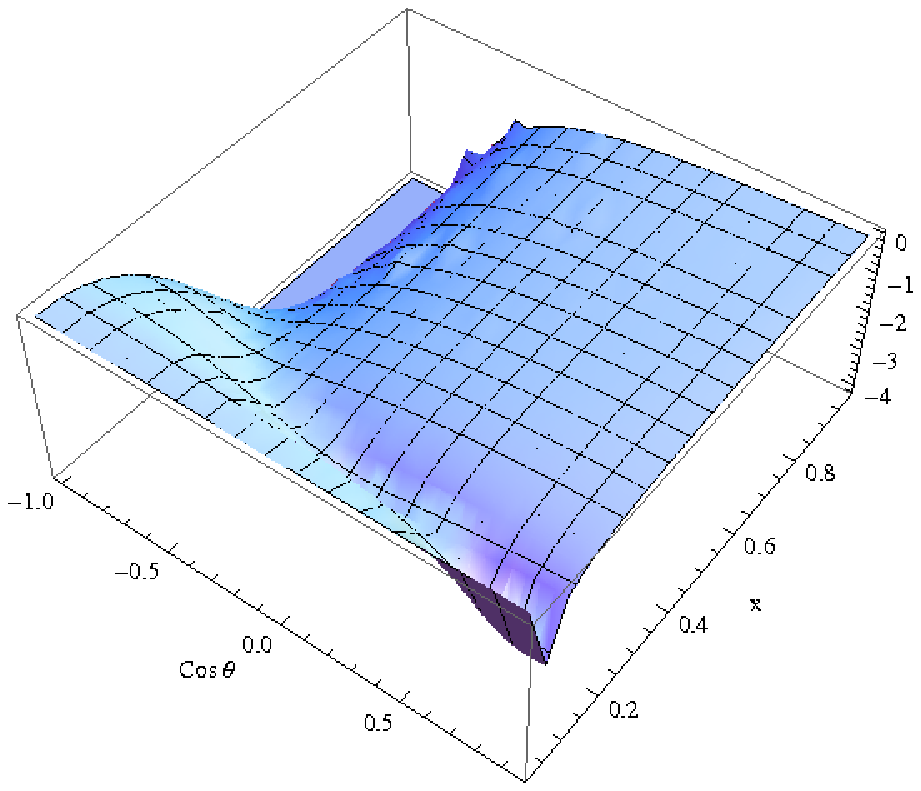,   width=6cm, angle=0}
\epsfig{file=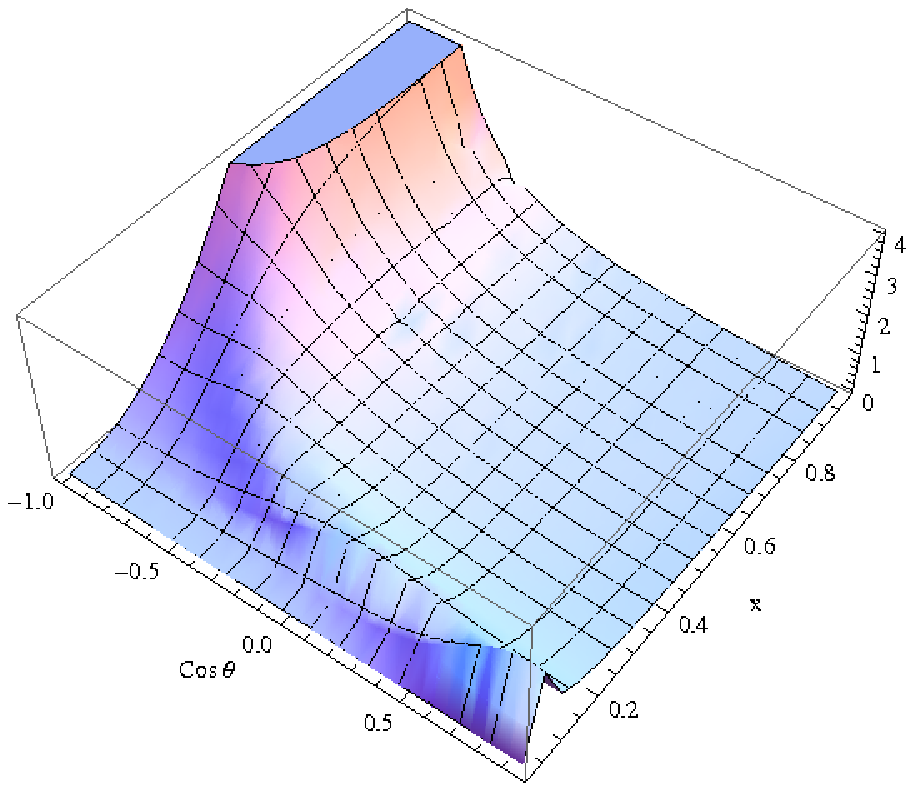, width=6cm, angle=0}
\caption{{3D plot showing deviation of the energy-angle distribution of secondary muons from the SM value for $\delta \kappa_\gamma=-0.072$ (left) and for $\delta \kappa_\gamma=+0.069$ (right) at $E_{cm}=500$ GeV.
}}
\label{fig:xct3Danom}
\end{center}
\end{figure}

Moving on to the energy and angle double-distribution of secondary muons, in Fig.~\ref{fig:dsigXCt} we plot the two-dimensional projections of this distribution at $E_{cm}=500$ GeV, where fixing the angle the distribution is studied by varying $E_\mu$. Figure on the left shows  case of SM. Clearly the energy distribution depends on what angle we are observing this. The peak of the distribution moves from lower values of $E_\mu$ to higher values as we go from smaller 
$\theta$ values to larger ones. Here we have a tool to probe new physics effects, which deviate from this behaviour. The case of anomalous coupling ($\delta \kappa_\gamma$) considered in this work behaves similar to that of the SM. Therefore for model discrimination we need to rely on quantitative analyses. At the same time, judicious choice of regions of $x-\theta$ plane can be made to improve sensitivity to new physics in this case. This is also clear from Fig.~\ref{fig:xct3D} and \ref{fig:xct3Danom}, where we can see that difference between cases with anomalous coupling and SM are larger in regions where $E_\mu$ is larger.
We may also mention that, it is possible that many new physics effects present even qualitatively different distributions. The energy-angle distribution will be more useful in such cases. A more complete analysis in this directions will be presented in a future work.

\section{Summary and Conclusions}

While, when considering the single $W$ production in $e\gamma$ collisions, angular distribution of secondary muons in the lab frame is easily constructed, and is readily available in the literature, such is not the case of its energy distribution. Here he have presented a semi-analytical way to explore the secondary lepton energy-angle distributions in 
$e\gamma\rightarrow \nu W$ with $W\rightarrow l\bar \nu$. The advantage of such an observable in analyzing the SM case and probing possible new physics effects is demonstrated. Variables being defined in the lab frame, can be directly used to apply experimental cuts. Many other observables like forward-backward asymmetry, etc. can be easily constructed from the distribution studied. As an example of new physics effect we considered the presence of CP-conserving dimension four anomalous $\gamma WW$ coupling. We studied the application of angular and energy cuts on the distributions, as well as studied the energy-angle distribution of the secondary muons, and conclude that judicious usage of observables discussed can probe the presence of new physics in single $W$ production in $e\gamma$ collisions more effectively.
Results presented are obtained with unpolarized beams. Since $W$ couples only to the left-handed electrons, fully left-polarized electron beams
along with unpolarized photon beams will only enhance (double) the statistics. At the same time, effect of photon beam polarization is unavoidable in $e\gamma$ collisions, and the Compton backscattered photons are not unpolarized. Also, we considered an ideal Compton backscattered spectrum \cite{photCollider} for our studies. To be more accurate, one need to consider a realistic photon spectrum including the non-linear effects, and the polarization of the photon beam.   
  \vskip 5mm
  \noindent
 {\large \bf Acknowledgement}: \\[2mm]
 PP's work is partly supported by a BRNS, DAE project.
 
\vskip 5mm
\noindent
 {\Large \bf Appendix: Photon luminosity distribution}\\[2mm]
The colliding photons in a realistic electron-photon collider does not have a fixed energy, rather the beam will have distribution of photons with energy varying over an allowed range (which depends on the initial electron and laser photon energies among other things). In such colliders, the cross section and other observables should, therefore, be properly folded with a luminosity distribution function to get the measurable quantities, as is done in Eq.~\ref{eqn:lumfold}.

 At ILC high energy, high luminosity photon beam is obtained by Compton backscattering of low energy, high intensity laser beam off high energy electron beam.  Ideal Compton backscattered photon spectrum is given by \cite{lumdist} 
 
  \begin{eqnarray}
           f_{\gamma/e}(x) &=& \frac{1}{D(\xi)}~\left[1-x+\frac{1}{1-x}-4~\frac{x}{\xi (1-x)}+4~\frac{x^2}{\xi^2(1-x)^2}\right] \nonumber \\
           D(\xi)&=&\left(1-\frac{4}{\xi}-\frac{8}{\xi^2}\right)~\ln\left(1+\xi\right)
           +\frac{1}{2}+\frac{8}{\xi}-\frac{1}{2(1+\xi)^2},
 \label{eqn:lumspec}
\end{eqnarray}
where $x=\frac{\omega}{E_e}$, with $E_e$ the energy of the initial electron and $\omega$  the energy of the scattered photon. $x$ thus gives the fraction of the electron energy carried by the scattered photon. 
Dependence of the distributions on the initial laser photon energy ($\omega_0$) 
comes through $\xi\approx \frac{4E_e\omega_0}{m_e^2}$, where $m_e$ is the electron mass. 
The maximum value of $x$ is $x_{max}=\frac{\xi}{1+\xi}$.  It is, but not possible to increase $\omega_0$ and $E_e $ to any value to get larger $x_{max}$. It is found that for $\xi$ beyond 
$\sim 4.8$, conversion efficiency drops down drastically due to $e^+e^-$ pair production between the laser photons and the backscattered photons, setting an absolute upper limit on $x\approx 0.83$. This value essentially means that with an electron beam of energy 
$E_e=250$ GeV, we can effectively go up to $\omega_0\approx 1.26$ eV . In a realistic collider, one need to also worry about the non-linear effects making the actual photon spectrum deviating from the ideal case in Eq.~\ref{eqn:lumspec}. For more details on photon collider one may refer to, for example, Ref.~\cite{lumdist, photCollider} and references therein.

\end{document}